# Graph statistics: An emerging discipline in non-Euclidean data analysis

Rongling Wu and Shing-Tung Yau

Beijing Key Laboratory of Topological Statistics and Applications for Complex Systems, Beijing Institute of Mathematical Sciences and Applications, Beijing 101408, China

Yau Mathematical Sciences Center, Tsinghua University, Beijing 100084, China

**Corresponding author:** Rongling Wu (ronglingwu@bimsa.cn)

## Abstract

The explosive growth of complex data has catalyzed the emergence of graph statistics as a fundamentally new discipline in data science. Unlike traditional statistics, which operates primarily within the comfortable confines of Euclidean spaces, graph statistics confronts the reality that modern data naturally organize themselves as dynamic networks composed of complex interconnections. In this article, we present an overview of graph statistics as an emerging discipline, tracing its theoretical foundations, methodological innovations, and transformative applications. We examine how the integration of evolutionary game theory, ecological niche theory, topological data analysis, and graph theory through quasi-dynamic nonlinear modeling has created a new norm of statistical thinking capable of analyzing non-Euclidean data. We show how graph statistics is poised to revolutionize fields ranging from quantitative genetics and systems biology to materials science and artificial intelligence, offering a principled framework for transforming big data into practical knowledge.

## Introduction

The twentieth century witnessed extraordinary advances in statistical methodology, from Fisher's foundational work on experimental design to the development of generalized linear models, mixed models, and survival analysis [1,2]. These methods share a common conceptual foundation: they treat observations as independent draws from some probability distribution, and they represent each observation as a point in a Euclidean space, a vector of measurements with a fixed dimension [3]. This Euclidean paradigm has served science remarkably well, enabling the analysis of agricultural field trials, clinical experiments, and industrial processes. Yet the data landscape of the twenty-first century has rendered this paradigm increasingly inadequate [4], because the systems that now demand statistical attention are fundamentally relational rather than independent.

Biological systems exemplify this relational nature most clearly. A gene does not function in isolation; it participates in regulatory circuits, metabolic pathways, and signaling networks [5]. A neuron does not compute independently; it integrates inputs from thousands of synaptic partners [6]. A species does not evolve alone; it coevolves within ecological communities shaped by competition, cooperation, and mutual dependence [7]. In each case, the essential properties of the system emerge not from the characteristics of individual components but from the patterns of connection between them. The relevant data are therefore not vectors of measurements but graphs of relationships. This observation carries a methodological imperative: to understand such systems statistically requires a shift from analyzing points to analyzing networks, from Euclidean geometry to graph topology, from reductionism to systems thinking.

This paradigm shift, from linear to nonlinear, from lines to curves, from individuals to systems, defines the intellectual program of graph statistics as developed by statistician R. Wu and mathematician S.-T. Yau [8-11]. The framework addresses its methodological imperative through two integrated pillars. The first is network reconstruction: the inference of relational structures from diverse data types, including gene expression profiles, microbial abundance measurements, phenotypic traits, and environmental covariates [12,13]. The second is topological mapping of network features: the identification of homological and geometric properties that characterize network organization and that may themselves be treated as phenotypes subject to genetic, environmental, or evolutionary influence [9-11,14].

The significance of this framework becomes apparent when contrasted with traditional approaches. Conventional statistical genetics, for instance, treats complex traits as scalar quantities, blood pressure, body mass index, and disease status, and seeks genetic variants that influence these quantities. But many traits are inherently relational. The structure of a gene regulatory network is a trait. The topology of a microbial interaction web is a trait. The connectivity pattern of a neural circuit is a trait. These network traits are not merely complicated scalar quantities; they represent qualitatively distinct kinds of phenotypes that require fundamentally different statistical tools. Graph statistics provides these tools, enabling researchers to pose questions that would be meaningless within the Euclidean paradigm: Which genetic variants determine whether a gene becomes a network hub? How does genetic variation influence the modular organization of cellular metabolism? What evolutionary forces shape the architecture of species interaction networks?

In this article, we introduce the conceptual and methodological foundations of graph statistics. We first clarify the distinction between Euclidean and non-Euclidean data, establishing why graph-structured data require fundamentally different statistical treatments. We then present evolutionary game theory as a principled foundation for understanding network formation, followed by multilayer modular sparse network reconstruction from high-dimensional data. The unified multi-task model integrates network estimation with variable selection on all agents, and the topological mapping framework connects network properties to underlying operation mechanisms. Furthermore, we demonstrate how graph statistics unlocks mechanistic interpretability within artificial intelligence architectures. Throughout, we emphasize how each component addresses specific challenges posed by relational data, and how the components together constitute a coherent methodology for analyzing non-Euclidean systems.

## What are Euclidean and non-Euclidean data?

The distinction between Euclidean and non-Euclidean data is foundational to understanding why graph statistics represents a genuine paradigm shift rather than a mere technical extension of existing methods. Euclidean data are those that can be naturally represented as points in a finite-dimensional vector space equipped with the standard inner product and metric structure [15,16]. Each observation consists of a fixed number of numerical measurements, and the distance between observations follows the Pythagorean theorem. Gene expression vectors, clinical biomarker profiles, and environmental measurement time series are canonical examples of Euclidean data. The statistical methods developed over the past century—regression, principal components analysis, discriminant analysis, clustering—are designed for and assume such data.

Non-Euclidean data violate one or more of the assumptions that define Euclidean structure [17]. Graph-structured data are the paradigmatic case. In a protein interaction network, the observations are not independent measurements but relationships between entities. The data consist of nodes (proteins) and edges (physical interactions), and the fundamental unit of analysis is the connection pattern rather than the node attribute. The space of all such networks is not a vector space. There is no natural way to add two networks or to multiply a network by a scalar. The dimension is not fixed: different biological samples may yield networks with different numbers of nodes and edges. And the distance between networks is not given by any inner product; the most natural metrics, such as the graph edit distance, are combinatorial quantities that cannot be expressed in terms of vector differences [18,19].

The non-Euclidean nature of graph data manifests in several specific ways that preclude direct application of classical methods. First, graphs lack a canonical ordering of nodes. The nodes of a gene co-expression network carry no intrinsic labels; any permutation of node identities produces the same network structure. This permutation invariance means that statistical procedures must yield identical conclusions regardless of arbitrary node labels, a constraint that has no analogue in Euclidean statistics. Second, graphs have variable dimensionality. One sample may yield a network of three hundred genes, another a network of three thousand. Standard multivariate methods require a fixed dimension and cannot accommodate such variation without ad hoc truncation or padding. Third, the natural operations on graphs are not linear. The composition of networks through union or intersection does not correspond to vector addition, and the scaling of edge weights does not correspond to scalar multiplication.

The consequences of non-Euclidean structure for statistical analysis are profound. In Euclidean spaces, the sample mean is the unique point that minimizes the sum of squared distances to the observations. In graph space, there is no unique notion of mean; different definitions (vertex-wise averages, graph edit distance minimizers, spectral embeddings) yield different objects with different properties. Similarly, variance, covariance, and correlation, central quantities in Euclidean statistics, lack straightforward analogues in graph space. Principal component analysis, which decomposes variation along orthogonal directions, requires a tangent space structure that graphs do not possess. These limitations are not mere inconveniences to be circumvented; they reflect genuine structural properties of relational data that any valid statistical approach must accommodate.

The recognition that graph data require specialized treatment has motivated the development of multiple methodological approaches. Spectral methods embed graphs into Euclidean spaces through eigen decomposition of the graph Laplacian, enabling the application of classical techniques at the cost of distorting the intrinsic geometry. Kernel methods define similarity measures between graphs that implicitly map them into high-dimensional feature spaces, though the mapping is typically intractable to compute explicitly. Graph neural networks learn parametric functions that operate directly on graph structure, achieving remarkable empirical success but often lacking theoretical interpretability. Wu and Yau's graph statistics framework [8-11] takes a distinct approach, integrating network reconstruction with dynamic models to connect the topological properties of networks to underlying operation mechanisms.

## Evolutionary game theory in action

The structure of complex networks is not arbitrary; in biology it reflects the outcome of evolutionary processes that shape interactions between genes, proteins, species, and individuals. Understanding why certain network architectures emerge and persist requires a theoretical framework that connects microscopic interaction rules to macroscopic topological patterns. Evolutionary game theory provides precisely such a framework, and its integration into prey-predator theory enables principled explanations of network formation.

Evolutionary game theory originated in the study of animal behavior, where Maynard Smith and Price introduced the concept of an evolutionarily stable strategy, a strategy that, when adopted by a population, cannot be invaded by any alternative strategy [20]. This framework has since been extended to network-structured populations, where the spatial arrangement of individuals determines who interacts with whom. In a network-structured population, each node represents an individual, edges represent interaction partners, and the fitness of each individual depends on the strategies of its neighbors. The evolutionary dynamics, how strategies spread through the network, are governed by the topology of the interaction graph [21,22].

The connection to graph statistics arises because the evolutionary dynamics themselves shape the network. If individuals can choose whom to interact with, or if interactions can be broken and reformed, then the network coevolves with the strategies on it. Beneficial partnerships are reinforced, detrimental ones dissolved, and the resulting equilibrium network structure reflects the underlying game-theoretic incentives. This coevolutionary perspective explains why certain network motifs, triangles, stars, bipartite structures, are overrepresented in biological networks [23]. Triangles emerge when mutual cooperation is stable, because the closed structure ensures that cooperative behavior is reciprocated. Stars emerge when central individuals can extract benefits from peripheral ones, creating hierarchical architectures. Bipartite structures emerge when two distinct types of entities interact exclusively with each other, as in mutualistic relationships between plants and pollinators.

In genetic networks, evolutionary game theory explains the emergence of hub genes, genes that interact with many others and occupy central positions in regulatory networks. From a game-theoretic perspective, a gene becomes a hub when its product participates in many beneficial interactions, each contributing to fitness. The stability of hub status depends on the cost of maintaining multiple interactions and the benefit derived from each. Genetic variants that alter these costs or benefits can shift a gene from peripheral to hub status, providing a mechanistic basis for network quantitative trait loci that influence network topology.

In microbial networks, evolutionary game theory illuminates the conditions under which cooperative interactions, such as cross-feeding, where one species consumes the metabolic byproducts of another, can persist despite the threat of exploitation by cheaters. The spatial structure of microbial communities, which can be represented as interaction networks, determines whether cooperators can form clusters that resist invasion. Host genetic variants that influence the spatial structure of the microbiome, through effects on gut morphology, immune surveillance, or nutrient availability, thereby shape the network of microbial interactions through game-theoretic mechanisms.

The integration of evolutionary game theory into graph statistics provides more than explanatory power; it generates testable predictions about how network structure should respond to perturbations. If a network motif is maintained by game-theoretic stability, then perturbing the payoffs should cause predictable changes in motif frequency. If hub status reflects evolutionary optimization, then introducing competing interactions should cause hub genes to lose centrality. These predictions can be tested through experimental manipulation and statistical comparison of network structures, linking the abstract mathematics of game theory to empirical biological inquiry.

The game-theoretic perspective also clarifies the relationship between network structure and function. In biological systems, function often emerges from the collective behavior of interacting components rather than from the properties of any single component. A metabolic network produces energy through the coordinated activity of many enzymes; a gene regulatory network specifies cell fate through the dynamic interplay of transcription factors. Evolutionary game theory shows how such collective functions can emerge from the selfish optimization of individual components, provided the interaction structure satisfies certain conditions. This insight guides the search for network features that are functionally significant: not the most highly expressed genes, but those whose network position maximizes their influence on system behavior.

## Multilayer modular sparse networks from high-dimensional data

The reconstruction of biological networks from high-dimensional data presents formidable statistical challenges. A typical transcriptomic dataset may measure the expression of twenty thousand genes in a few hundred samples. The number of possible gene pairs, potential network edges, far exceeds the number of observations, making the estimation problem ill-posed without additional structure. The multilayer modular sparse network approach addresses this challenge by integrating three key ideas: multilayer structure to capture different types of relationships, modularity to identify functionally coherent groups, and sparsity to select relevant connections from the vast space of possibilities.

The multilayer aspect recognizes that biological systems involve multiple types of relationships that operate on different scales and timescales. In a host-parasite system, one layer may represent gene co-expression within the host, another layer may represent gene co-expression within the parasite, and a third layer may represent cross-species interactions. Each layer has its own statistical properties, different variances, different correlation structures, different noise levels, and the connections between layers carry information about how the systems interact. Modeling these layers separately would ignore critical cross-layer dependencies; modeling them as a single homogeneous network would obscure important structural differences. The multilayer framework strikes a balance, allowing each layer to have distinct characteristics while modeling inter-layer connections explicitly.

Modularity addresses the observation that biological networks are not random graphs but organized into communities or modules, groups of nodes that are densely connected internally but sparsely connected to other groups [24]. In gene co-expression networks, modules often correspond to functional pathways: one module may contain genes involved in cell cycle regulation, other genes involved in immune response, a third genes involved in metabolic processing. Identifying these modules is both a statistical challenge and a biological objective, as modules provide a natural level of abstraction for understanding network function. The modular structure also aids statistical estimation: by restricting connections to within-module and between-module patterns, the effective number of parameters is reduced, mitigating the curse of dimensionality.

Sparsity is the principle that most possible connections in a biological network do not exist or are negligible [25]. Of the hundreds of millions of possible gene pairs, only a small fraction is functionally relevant. Enforcing sparsity in network estimation prevents overfitting, improves interpretability, and reflects biological reality. Various sparsity-inducing techniques have been developed, including regularization methods that penalize the number or magnitude of edges, thresholding procedures that retain only the strongest observed correlations, and Bayesian methods that place sparsity-promoting priors on edge probabilities [26,27]. The choice of sparsity method involves tradeoffs between computational tractability, statistical consistency, and biological plausibility.

The integration of multilayer structure, modularity, and sparsity enables network reconstruction from high-dimensional data with strong biological interpretability. The estimation procedure typically proceeds in stages. First, variables are clustered into preliminary modules based on marginal associations or prior biological knowledge. Second, within each module and between module pairs, relevant connections are selected through sparse regression or penalized likelihood methods. Third, the multilayer structure is incorporated by estimating cross-layer connections that respect the modular organization. The result is a network that captures the essential relational structure of the data without being overwhelmed by noise or dimensionality.

Functional clustering and variable selection are closely intertwined in this framework [28-31]. Clustering identifies groups of variables that behave similarly, reducing the effective dimensionality and suggesting functional hypotheses. Variable selection identifies, within each cluster, the specific variables that drive the observed associations, distinguishing true regulatory relationships from spurious correlations. Together, these operations transform raw high-dimensional data into structured network representations that can be interrogated for biological meaning.

The application to microbial networks illustrates these principles. Microbiome data consist of abundance measurements for hundreds of bacterial taxa across samples. The multilayer structure may distinguish taxa that co-occur within the same body site from taxa that co-occur across sites. Modularity identifies microbial communities, groups of taxa that tend to appear together, potentially reflecting metabolic interdependencies or shared environmental preferences. Sparsity ensures that only the strongest and most consistent associations are retained, producing a network that highlights key ecological relationships rather than overwhelming detail. The resulting network can then be mapped to host genetic variation, identifying genetic variants that influence microbial community structure.

## Multi-task learning of idopNet

Informative, dynamic, omnidirectional, and personalized networks (idopNet) represent a cutting-edge paradigm in modern system architecture. The reconstruction of idopNet rests on a mathematical foundation that decomposes the dynamic behavior of each system agent $j$ ($j = 1, \ldots, m$) into two fundamental components [8,12], expressed as

$$\frac{dy_j}{dt} = Q_j(y_j) + \sum_{j'=1, j' \neq j}^{m} Q_{jj'}(y_{j'})$$

where the first term at the right-hand side is an independent component reflecting agent $j$'s intrinsic capacity to function without external influence and the second comprises dependent components arising from the cumulative influence of other agents $j'$ ($j' = 1, \ldots, j-1, j+1, \ldots, m$) on this agent. It can be seen that these components are formalized through a system of mixed ordinary differential equations (MODE), in which each equation governs the temporal evolution of a single agent's state. The independent component appears as a function of the agent's own state, while the dependent components appear as functions of the states of interacting agents. This decomposition naturally maps onto a graph representation: agents become nodes, and their interactions become edges weighted by the strength, sign, and direction of influence.

Two critical challenges arise in idopNet reconstruction from this framework. The first is variable selection. Empirical systems typically involve many more potential interactions than available observations, making the estimation problem ill-posed without additional structure. Variable selection imposes sparsity by identifying the subset of interactions that are genuinely informative, filtering out noise and irrelevant associations. The second challenge is solving the MODE in a manner that accurately encodes the independent and dependent components into nodes and edges. Conventional approaches have treated these two challenges separately, and have further handled each variable individually, ignoring the interdependence among variables that is the defining feature of complex systems.

Multi-task learning addresses these limitations by integrating variable selection and equation solving into a unified matrix framework [32]. Rather than selecting variables and estimating interactions for each agent independently, multi-task learning performs these operations simultaneously across all agents, exploiting the shared structure of the interaction network. The framework is built on the matrix representation of nonlinear additive regression models derived from the system of mixed ordinary differential equations. Nonlinear additive regression preserves the inherent nonlinearity of complex systems while enabling computationally tractable estimation through additive decomposition.

A practical obstacle complicates this theoretical framework. The MODE are built on a temporal scale, yet biological and ecological studies rarely collect high-density time series data. The cost and logistical difficulty of repeated sampling often restrict researchers to static snapshots, single measurements across many samples rather than trajectories across time. To overcome this limitation, ecological niche theory is introduced to extract dynamic information from static data [8]. The allometric scaling law, which relates organismal traits to body size across species, provides a principled basis for inferring dynamic trajectories from cross-sectional observations [33]. By treating samples as points along an inferred temporal or ecological gradient, a system of quasi-dynamic MODE (qdMODE) is constructed. These quasi-dynamic equations reveal variable trajectories over samples rather than over time, preserving the mathematical structure of the dynamic framework while accommodating realistic data constraints. The theoretical properties of this quasi-dynamic extension have been established and validated [34]. Figure 1 outlines the core advantages of idopNet over traditional networks found in recent literature. By leveraging evolutionary game theory, the proposed architecture dynamically maps the independent effects (node) and dependent interactions (edges) within complex systems.

Within this matrix framework, an iterative double-sparse hard thresholding algorithm implements the multi-task learning objective [35]. The algorithm enforces two coupled sparsity constraints simultaneously. Group-level sparsity identifies modules or communities of agents that interact densely among themselves but sparsely with other groups. Elementwise sparsity selects specific interaction edges within and between these modules. The coupling of these constraints ensures that the modular structure and the fine-grained interaction pattern are mutually consistent, avoiding the contradictory estimates that arise when modules and edges are selected independently.

The multi-task learning formulation operates on nonlinear feature representations that capture complex, non-additive relationships between agents. Unlike linear approximations that assume proportional effects and independent contributions, these representations accommodate saturation effects, threshold behaviors, and synergistic interactions characteristic of real biological systems. The matrix form enables scalable computation through efficient linear algebra operations, while the multi-task structure ensures that sparsity patterns are coherent across the entire network.

This unified approach transcends conventional network reconstruction methods in several respects. Existing practice typically assumes pairwise, symmetric, or static interactions, and relies on linear correlation or mutual information that cannot capture nonlinear dependencies. Multi-task learning-based idopNet capture omnidirectional interacting links that may be directed, weighted, and higher-order. They accommodate nonlinearities in both individual agent dynamics and inter-agent coupling. They reveal emergent properties, collective behaviors arising from interaction patterns that cannot be predicted from individual agent properties alone. These capabilities enable idopNet to approximate the true complexity of complex systems to a substantially larger extent than previous methods.

The practical utility of this framework is demonstrated through application to gene regulatory network reconstruction for the malaria parasite *Plasmodium falciparum* [32]. Using transcriptional data from infection time courses, the multi-task learning model identifies previously unknown regulatory roles of several genes in mediating malaria infection. These findings illustrate how the integration of ecological niche theory, evolutionary game dynamics, and machine learning within a non-Euclidean statistical framework can generate novel biological insights from high-dimensional genomic data. The model provides a principled approach for analyzing, modeling, and interpreting complex relational data in spaces where topology rather than coordinates carries the essential information.

## GLMY dissection of idopNet

The reconstruction of a complex network is not an end in itself but a means to understand how network features shape natural function. The GLMY framework, pioneered by S.-T. Yau and his team, aims to identify topological features of digraphs that govern system behavior [36]. These features, encompassing connectivity patterns, cycle structures, and hierarchical organization, provide a bridge between abstract network topology and concrete biological or physical processes. By mapping how these topological properties influence functional outcomes, the framework enables researchers to move beyond descriptive network analysis toward predictive understanding of why certain architectures persist under selective pressure and how they might be manipulated for therapeutic or engineering purposes.

The GLMY theory is grounded in a deep mathematical principle: the topology of a directed graph encodes essential information about the dynamical systems that operate upon it [37]. The framework decomposes complex digraphs into fundamental topological invariants that remain robust under continuous deformation. These invariants include source-sink structures that govern information or material flow, feedback cycles that enable self-regulation and homeostasis, and stratified hierarchies that organize system components into functional levels. The mathematical foundation draws upon algebraic topology, particularly the theory of persistent homology adapted to directed structures, and differential geometry, where curvature properties of network paths determine stability and convergence of associated dynamical processes. The GLMY framework thus provides a principled language for connecting static network architecture to dynamic system behavior, a connection that is essential for understanding how form determines function in complex systems.

The application of GLMY dissection to metabolic networks illustrates its biological utility [9]. Cellular metabolism is organized as a directed graph in which nodes represent metabolites and edges represent enzymatic reactions. The topological features identified by GLMY analysis, source metabolites that enter the network from external uptake, sink metabolites that are secreted or consumed in biomass production, and cyclic pathways that regenerate cofactors, directly correspond to functional modules recognized by biochemists. However, GLMY dissection goes beyond conventional pathway annotation by revealing higher-order topological constraints that limit metabolic flux and determine robustness to genetic perturbation. For instance, the presence of short feedback cycles in a metabolic network predicts the capacity for rapid allosteric regulation, while the hierarchical depth of the network reflects the evolutionary layering of ancient and recently acquired metabolic capabilities. By identifying these topological features, GLMY analysis enables the prediction of metabolic phenotypes from network structure alone, guiding metabolic engineering strategies for the production of biofuels, pharmaceuticals, and commodity chemicals.

Microbial interaction networks present a distinct but equally powerful application domain [10,14]. In these networks, nodes represent microbial species or strains, and directed edges represent cross-feeding, competition, or signaling interactions. The GLMY framework identifies topological features that determine community stability and function: source species that introduce carbon or nitrogen into the community, sink species that consume fermentation products and prevent toxic accumulation, and cyclic interaction motifs that mediate cooperative feedbacks. These topological properties explain why certain microbial consortia assemble predictably from environmental inocula while others remain unstable and functionally inconsistent. The hierarchical organization revealed by GLMY dissection corresponds to trophic levels in microbial food webs, with implications for community engineering in bioremediation, agriculture, and human health. Understanding how topological features govern microbial community function enables the rational design of synthetic consortia for targeted biotechnological applications.

Beyond biology, the GLMY framework finds natural application in materials science, where the topological features of atomic or molecular interaction networks determine macroscopic material properties [38]. In crystalline materials, the directed graph of atomic bonds exhibits topological invariants that predict mechanical strength, thermal conductivity, and electronic band structure. In amorphous materials and glasses, the absence of long-range order makes conventional crystallographic analysis inapplicable, yet GLMY topological features, local source-sink configurations, persistent cycles in the bond network, and hierarchical clustering of atomic environments, provide predictive descriptors of rigidity transitions and relaxation dynamics. In soft materials and polymers, the directed topology of crosslink networks determines viscoelastic response and self-healing capacity. The GLMY framework thus provides a unified mathematical language for analyzing network-function relationships across biological and physical systems, revealing common topological principles that govern complex behavior regardless of specific domain details.

**How graph statistics enables graph neural networks**

Graph statistics enables artificial intelligence (AI) by providing mathematical foundations, mechanistic interpretability, and geometric constraints for graph neural networks (GNNs). While deep learning models achieve massive empirical success across relational datasets, current GNN architectures face severe interpretability limitations [39]. They operate primarily as predictive black boxes, mapping input nodes to target labels without revealing the operational laws governing those systems [40]. The graph statistics framework solves this core challenge. It integrates quasi-dynamic nonlinear modeling to transform how AI treats non-Euclidean relational structures. Unlike traditional data points, networks reside on directed, signed, and weighted interdependence. Classical vector spaces cannot add networks or multiply them by scalars. Graph metrics use combinatorial quantities instead, making standard linear operations invalid. By unifying algebraic topology, differential geometry, and machine learning, this framework establishes principled data scaling laws that bridge empirical success with rigorous theory.

This synthesis of disciplines directly confronts the core structural limitations of modern AI. In classical machine learning, data is represented as static vectors in flat Euclidean spaces where distances follow the Pythagorean theorem. However, modern relational systems, such as gene regulatory circuits, synaptic pathways, and species interaction webs, are fundamentally non-Euclidean. While GNNs capture graph structures through localized message passing, their reliance on permutation-invariant aggregation means that they lack a rigid spatial coordinate system to anchor node orientations [41]. Graph statistics rectifies this by offering a formal topological framework. It proves that network features can be treated as distinct phenotypes governed by precise evolutionary and environmental pressures, moving AI from arbitrary pattern recognition to rule-based structural discovery.

Furthermore, graph statistics provides a solution to the variable dimensionality and permutation invariance issues that plague neural networks. Because graphs lack a canonical ordering of nodes, any arbitrary shuffling of labels alters the input matrix without changing the underlying network architecture [42]. Graph statistics provides the mathematical invariants needed to ensure that AI algorithms yield identical conclusions regardless of node indexing. Similarly, while deep learning typically struggles with inputs of varying sizes [43], the graph statistics framework easily accommodates networks of fluctuating node and edge counts without relying on ad-hoc truncation or padding.

By incorporating evolutionary game theory and ecological niche theory, this methodology also infuses predictive models with a layer of mechanistic causality [44]. Instead of blindly optimizing weights, GNNs can leverage graph statistics to model network formation as a system of qdMODE. This allows AI to decompose system behavior into independent intrinsic capacities and dependent interactive components. Rather than treating connections as simple symmetric weights, AI can now learn omnidirectional, directed, and higher-order links. Consequently, the resulting architectures do not merely predict outcomes; they explain why specific network configurations exist, how hubs form, and how the system will respond to physical or biological perturbations.

## Concluding remarks

The graph statistics framework developed jointly by statisticians and mathematicians represents a comprehensive response to the challenge of analyzing non-Euclidean data in the era of big data. By shifting the statistical paradigm from linear to nonlinear, from isolated observations to interconnected systems, this framework enables researchers to model biological complexity at the level where function actually emerges. The integration of network reconstruction, evolutionary game theory, multilayer modular sparse estimation, unified multi-task modeling, and genetic mapping of network features provides a coherent methodology that addresses the full pipeline from raw data to biological insight.

Looking forward, the convergence of graph statistics with other theoretical frameworks promises further advances. The Yau-Yau filtering theory, which provides a finite-dimensional coordinate approach to nonlinear filtering, offers natural connections to dynamic network models where latent states evolve continuously while observations arrive discretely [45-48]. Curvature theory, which characterizes the geometric properties of network spaces, illuminates the constraints on information flow and learning in graph neural networks [49]. Mean field theory, which approximates the behavior of large interacting systems through averaged equations, provides scalable tools for analyzing networks with millions of nodes [50]. The unification of these perspectives, graph topology, stochastic dynamics, differential geometry, and statistical physics, defines the frontier of non-Euclidean data analysis.

The ultimate measure of any statistical framework is its capacity to generate scientific understanding that would otherwise remain inaccessible. Graph statistics meets this standard by revealing how network features shape the structure and function of complex systems, how evolutionary forces mold the topology of interaction networks, and how collective function emerges from relational structure. As data continue to proliferate across physical, biological, and social scales, from molecular interactions to ecosystems to economic complexity, the need for methods that respect the non-Euclidean nature of reality will only intensify. The paradigm shift embodied in graph statistics is not merely a methodological convenience but a necessary adaptation to the structure of the natural world.

**Acknowledgements:** The authors sincerely thank Dr. Zhen Li for his constructive reviews and meticulous reading of the manuscript. This work was supported in part by the Shanghai Institute of Mathematics and Interdisciplinary Sciences (grant number: SIMIS-ID-2024-WN).

**Competing interests:** The authors declare no competing interest.

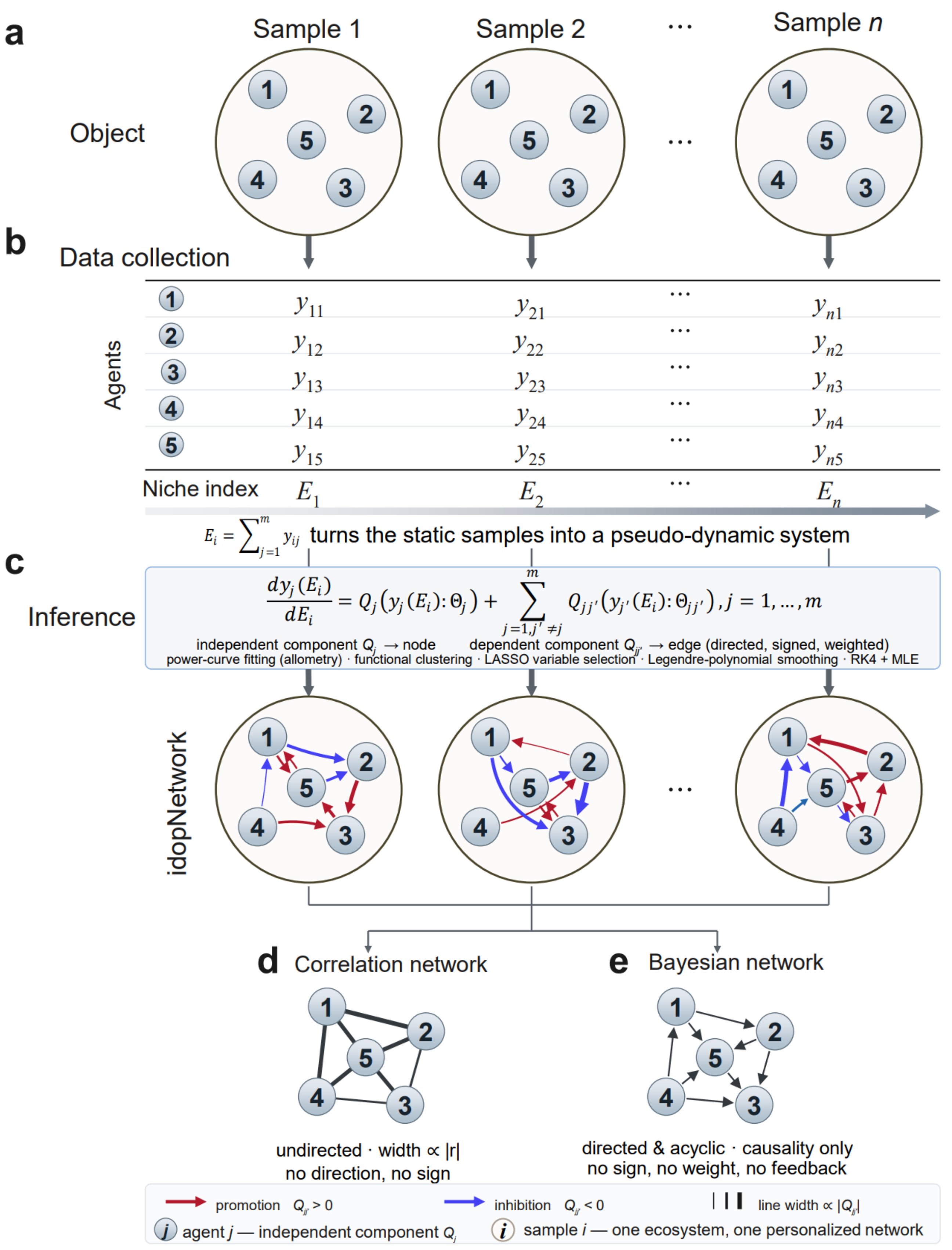

**Figure 1. Reconstructing idopNetwork: From static data to causative networks.**
(a) **Sample collection:** A panel of $n$ studied objects are sampled from 1 to $n$.

(**b**) **Agent-based data profiling:** Data collection from a set of agents (e.g., five) that characterize specific object features within a complex system.
(**c**) **idopNetwork inference:** Individualized networks are reconstructed from the sample panel using qdMODE. These causative networks are uniquely informative (bidirectional, signed, and weighted), dynamic (tracing interactions across spatiotemporal scales), omnidirectional (capturing all underlying interactions), and personalized (varying from sample to sample). Within the idopNetwork, red and blue arrows indicate promotional and repressional regulations, respectively, with line thickness proportional to interaction strength.
**Traditional approaches:** Conventional methods infer a single, static network pooled from the entire panel, yielding either (**d**) correlation-based networks (weighted but nondirectional) or (**e**) Bayesian networks (directional but unsigned).